\documentclass[%
reprint,
superscriptaddress,
 amsmath,amssymb,
 aps,
floatfix,
]{revtex4-1}

\usepackage{graphicx}% Include figure files
\usepackage{dcolumn}% Align table columns on decimal point
\usepackage{bm}% bold math
\usepackage{float}
\usepackage{mathptmx}
\usepackage[utf8]{inputenc}
\usepackage[T1]{fontenc}
\usepackage{marvosym}
\usepackage{amsmath}
\usepackage{amsbsy}
\usepackage{appendix}
\usepackage{hyperref}
\usepackage{textcomp}
\hypersetup{
    colorlinks=true,
    linkcolor=black,
    citecolor=blue,
    urlcolor=blue,
}
\usepackage{ragged2e}
\begin{document}

\preprint{APS/123-QED}

\title{%\color{blue} Optimization of terahertz emission from a full Heusler based spintronic emitter\\
%or\\
\textbf{\color{blue} Co$_2$FeAl full Heusler compound based spintronic terahertz emitter}}

\author{Rahul Gupta}
   \email{rahul.gupta@angstrom.uu.se; rahulguptaphy@gmail.com}
  \affiliation{
 Department of Materials Science and Engineering, Uppsala University, Box 35, SE-751 03 Uppsala, Sweden
}   
\author{Sajid Husain}
\affiliation{
Department of Materials Science and Engineering, Uppsala University, Box 35, SE-751 03 Uppsala, Sweden
}

\author{Ankit Kumar}
\affiliation{
Department of Materials Science and Engineering, Uppsala University, Box 35, SE-751 03 Uppsala, Sweden
}
\author{Rimantas Brucas}
\affiliation{
Department of Materials Science and Engineering, Uppsala University, Box 35, SE-751 03 Uppsala, Sweden
}
\affiliation{Ångström Microstructure Laboratory, Uppsala University, Box 35, SE-751 03 Uppsala, Sweden}
\author{Anders Rydberg}
\affiliation{
Department of Materials Science and Engineering, Uppsala University, Box 35, SE-751 03 Uppsala, Sweden
}
\affiliation{Department of Physics and Astronomy, FREIA, Box 516, SE-75120 Uppsala, Sweden}

\author{Peter Svedlindh}%
\email{peter.svedlindh@angstrom.uu.se}
\affiliation{
Department of Materials Science and Engineering, Uppsala University, Box 35, SE-751 03 Uppsala, Sweden
}

\date{\today}
 
\begin{abstract}
To achieve a large terahertz (THz) amplitude from a spintronic THz emitter (STE), materials with 100\% spin polarisation such as Co-based Heusler compounds as the ferromagnetic layer are required. However, these compounds are known to loose their half-metallicity in the ultrathin film regime, as it is difficult to achieve L2$_1$ ordering, which has become a bottleneck for the film growth. Here, the successful deposition using room temperature DC sputtering of the L2$_1$ and B2 ordered phases of the Co$_2$FeAl full Heusler compound is reported. Co$_2$FeAl is used as ferromagnetic layer together with highly orientated Pt as non-ferromagnetic layer in the Co$_2$FeAl/Pt STE, where an MgO(10 nm) seed layer plays an important role to achieve the L2$_1$ and B2 ordering of Co$_2$FeAl. The generation of THz radiation in the CFA/Pt STE is presented, which has a bandwidth in the range of 0.1-4 THz. The THz electric field amplitude is optimized with respect to thickness, orientation, and growth parameters using a thickness dependent model considering the optically induced spin current, superdiffusive spin current, inverse spin Hall effect and the  attenuation of THz radiation in the layers. This study, based on the full Heusler Co$_2$FeAl compound opens up a plethora possibilities in STE research involving full Heusler compounds.
\end{abstract}

\maketitle

%\section{\label{sec:level1}introduction\protect\\}

Due to several applications of terahertz (THz) radiation in fundamental and applied sciences \cite{jin2015accessing,hangyo2005terahertz,yang2016biomedical,zeitler2007terahertz,parrott2011terahertz,zeitler2007terahertz}, it is of eminent interest to develop broadband, powerful and low-cost THz radiation sources. The THz radiation lies in the frequency range from 0.1 to 30 THz (3.34 cm$^{-1}$ $-$ 1000.69 cm$^{-1}$) and  coincides with the energy scale of many collective excitations in the materials \cite{Zhang2020terahertzmagneto,Peter2011terahertz}. For example, the resonance frequency of second-order magnetization dynamics in ferromagnetic materials, governed by the inertia of the magnetic moment, lies in THz frequency range  \cite{PhysRevB.83.020410,neeraj2020inertial}. For decades, many efforts have been made towards the development of broadband, powerful and low cost THz emitters. Most of the  THz emitters nowadays are based on photo-carrier excitation of different materials using femtosecond laser pulses \cite{Prieto_2020}. Recently, it has been experimentally demonstrated that THz radiation can also be generated in ferromagnetic (FM) and non-ferromagnetic (NFM) magnetic heterostructures, where apart from the charge of the electron, the spin degree of freedom is utilized to generate the THz radiation \cite{kampfrath2013terahertz,Seifert2016efficient}. These magnetic heterostructures are generally referred to as spintronic THz emitters (STEs). In fact, for some of the STEs, such as CoFeB/(Pt,W,Pd) and Fe/Pt, a THz bandwidth in the range of 0.3-30 THz have been reported \cite{Seifert2016efficient,yang2016powerful}. However, it should be noticed that the the experimentally observed bandwidth depends on the pulse width of the femtosecond laser and the detector used in the terahertz-time domain (THz-TDS) set-up \cite{Seifert2016efficient}.

\begin{figure*}[t]
    \centering
    \includegraphics[width=18cm]{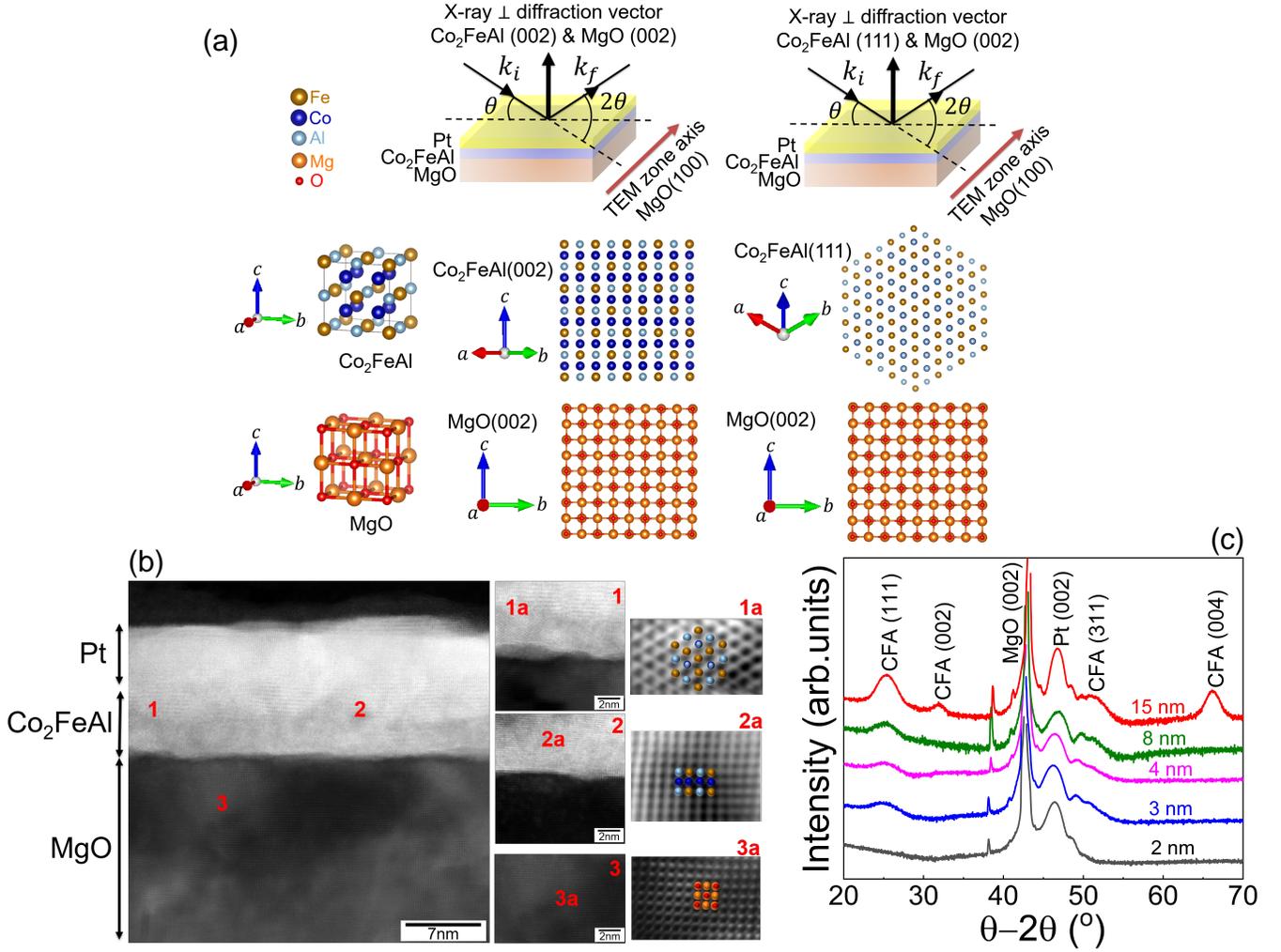}
    \caption{(a) In the first row, the colors of atoms, the perpendicular diffraction vector ($k_f-k_i$) for CFA(002)/MgO(002) Bragg peaks, and for CFA(111)/MgO(002) Bragg peaks in X-ray diffraction geometry are shown from the left to right, respectively. In the second row, the unit cell of CFA, the atomic orientations corresponding to CFA(002) and CFA(111) Bragg peaks are shown from left to right, respectively. In the third row, the unit cell of MgO, the atomic orientations corresponding to MgO(002), and MgO(002) Bragg peaks are shown from left to right, respectively. In this model, the cross-sectional TEM zone axis is parallel to MgO [100] for the atomic arrangements of CFA(111), CFA(002), and MgO(002). 
    (b) Cross-sectional TEM image of Pt(6 nm)/CFA(8 nm)/MgO(10 nm)/MgO,  where 1, 2 and 3 indicate regions  with L2$_1$ CFA phase, B2 CFA phase and MgO phase, respectively. (c) X-ray diffraction Bragg peaks for different CFA thickness with Pt(6 nm).}
    \label{fig:my_label}
\end{figure*}

In a THz-TDS set-up, a femtosecond laser pulse creates a non-equilibrium spin polarization in the FM layer, which superdiffuses through the FM/NFM interface into the NFM layer. The spin current is converted to a charge current in the NFM layer via the inverse spin Hall effect, which generates the electromagnetic radiation in the THz range. The output signal of the STE thus depends in three fundamental processes; generation of optically induced spin current in the FM (ultrafast dynamics in femtosecond region), spin-dependent interfacial transport across the FM/NFM interface, and the spin to charge conversion efficiency of the NFM layer due to spin-orbit coupling (SOC). The ideal STE consists of a $\sim$100\% spin polarised FM layer, an NFM layer with a large SOC (e.g., Pt, W, etc.)\cite{husainreview}  and a sharp interface between the layers. Keeping the aforementioned factors in mind, a search for a $\sim$100\% spin polarised FM material and large SOC NFM layer is required. Co-based full Heusler compounds are known to exhibit large spin polarizations ($P$);  e.g. Co$_2$MnSi ($P=59-66$\%), Co$_2$MnGe ($P=70$\%), and Co$_2$FeAl ($P=86$\%) \cite{PhysRevX.2.041008,muller2009spin}. Among the full Heusler compounds, Co$_2$FeAl (CFA) belongs to the $Fm\Tilde{3}m$ space group, exhibits half-metallicity and large ($\sim$86\%) spin polarization, which makes CFA a suitable candidate not only in spin-based electronic devices but also for STEs.

CFA can exhibit three types of ordering; fully ordered phase (L2$_1$), partially ordered phase (B2) and disordered phase (A2). The L2$_1$ phase, being half metallic, is shown in Figure 1. The B2 phase is formed when Fe and Al randomly share their lattice sites, whereas when Co, Fe, and Al are randomly distributed on the lattice sites it corresponds to the A2 phase. For the L2$_1$ phase, half-metallicity results from hybridization of the transition-metal element $d$-orbitals forming bonding (2e$_g$ and 3t$_{2g}$) and non-bonding states (2e$_u$ and 3t$_{1u}$). The half-metallicity emerges from the separation of non-bonding hybrids of Co as it cannot hybridize with the $d$-orbitals of Fe \cite{PhysRevB.66.174429,fecher2007substituting,PhysRevB.87.024420}. Depending upon the chemical ordering between Co, Fe and Al in CFA, the half-metallicity  may be reduced due to modification of the hybridization between the $d$-orbitals of Co and Fe \cite{fecher2007substituting,PhysRevB.87.024420}. The atomic ordering depends on thin film growth parameters such as underlying substrate, buffer layer, growth temperature etc.

To have a large amplitude of the THz-electric field, the thickness of the FM and NFM layers needs to be in the nanometer range. However, it is difficult to achieve the L2$_1$ ordered phase of CFA in ultralow thickness regime ($\sim$3-5 nm). Belmeguenai \textit{et al.} \cite{PhysRevB.87.184431} reported that B2 ordering of CFA can be achieved for 20 nm thickness, which is too large for CFA based STEs. Husain \textit{et al.} \cite{husain2016growth,husain2019multi} also reported B2 ordering of CFA deposited on MgO and Si substrates for the  relatively larger thickness regime ($\sim$50 nm). Moreover, L2$_1$ ordering has been achieved for Co$_2$FeSi Heusler compounds deposited on MgO, SrTiO$_3$, and MgAl$_2$O$_4$ substrates having relatively larger thickness ($\sim$25 nm) \cite{kudo2019great}.
X. Zhang \textit{et al.} reported that the spin polarization varied between 50-60\% varying the thickness of B2 ordered CFA films in the range of 5-20 unit cells ($\sim$2.8-12 nm) \cite{zhang2018direct}. 

\begin{figure*}[t]
    \centering
    \includegraphics[width=20cm]{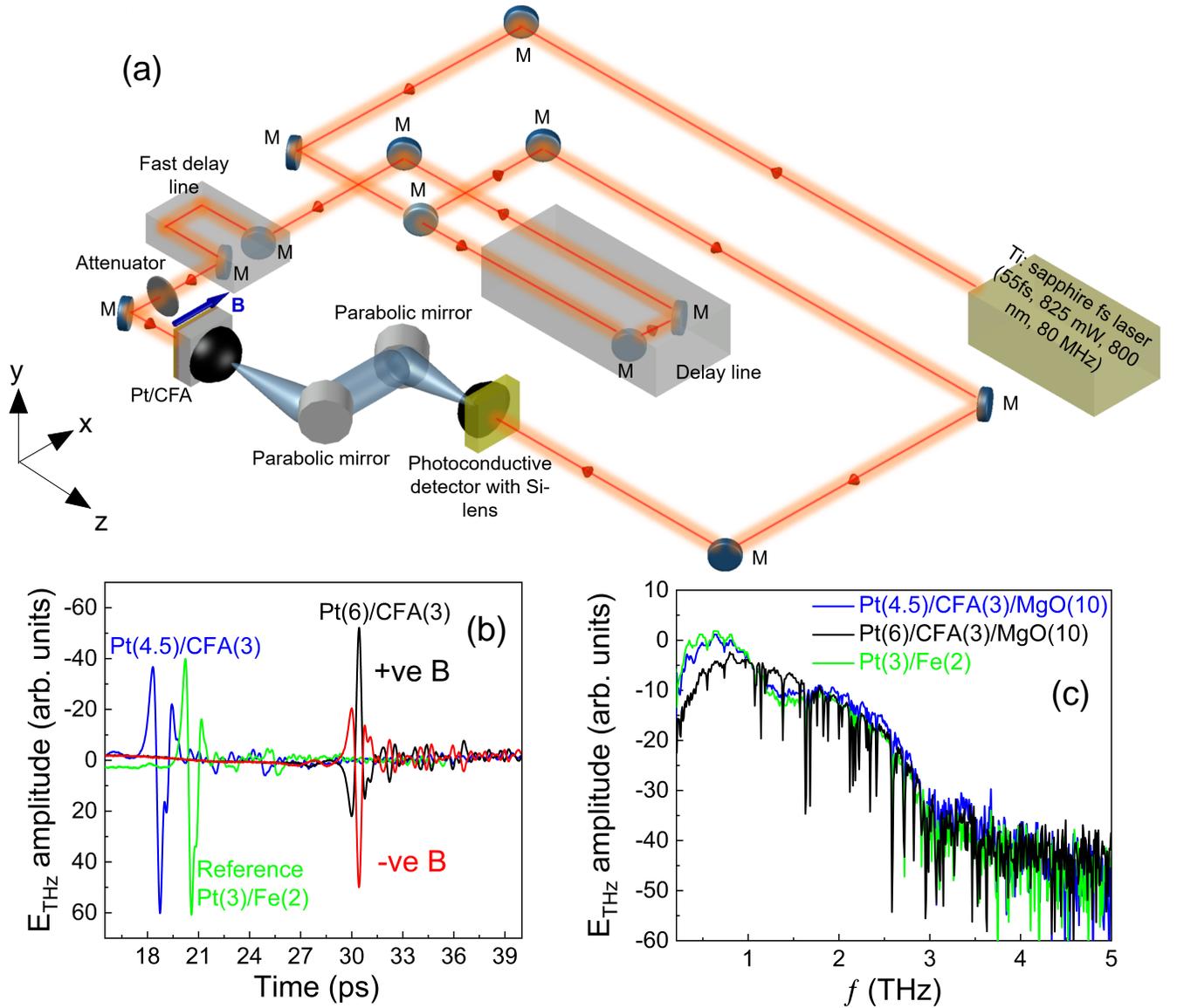}
    \caption{(a) A schematic of the experimental THz-TDS set-up. (b) Typical spectrum of THz electric field amplitude in time-domain for the Pt(4.5)/CFA(3)/MgO(10)/MgO (blue color), Pt(3)/Fe(2)/MgO (green color) and Pt(6)/CFA(3)/MgO(10)/MgO STEs (black/red color). (c) Fast Fourier transform of time-domain signal for the Pt(4.5)/CFA(3)/MgO(10)/MgO (blue color), Pt(3)/Fe(2)/MgO (green color) and Pt(6)/CFA(3)/MgO(10)/MgO STEs (black color).}
    \label{fig:my_label}
\end{figure*}
In this work, we have prepared highly oriented CFA/Pt STEs on MgO and Si substrates using room temperature DC magnetron sputtering, achieving mixed  L2$_1$ and B2 ordering of CFA down to 3 nm thickness. Because of mixed L2$_1$ and B2 ordering, which will be discussed in the forthcoming section, we expect more than 50-60\% spin polarization in our CFA thin films. To achieve the L2$_1$ and B2 ordering, a MgO 10 nm seed layer was used on MgO and Si substrates (cf. supplementary information section I). To confirm the L2$_1$ and B2 phases of CFA and the orientation of the Pt layer in our CFA/Pt STEs, we have performed X-ray diffraction measurements in the Bragg-Brentano geometry using a Cu K$_\alpha$ source. The presence of CFA(111), CFA(002) and Pt(002) diffraction peaks indicate that the CFA film exhibits a mixed L2$_1$ and B2 phase and that Pt is highly orientated following the CFA(002) orientation. However, the CFA(002) and CFA(004) Bragg peaks do not appear below 15 nm CFA thickness as shown in Figure 1(c). This may be explained by the X-ray source not having enough intensity to resolve these peaks. To confirm our supposition, we utilized  cross-sectional transmission electron microscopy (TEM) to visualize the atomic orientation of CFA with respect to the MgO-substrate for the Pt(6 nm)/CFA(8 nm)/MgO(10 nm)/MgO sample. As expected, two atomic orientations,  CFA(111) and CFA(002), can be identified in the TEM images. On the one hand, when we look at the structure along the MgO [100] substrate direction in cross-sectional TEM, MgO and CFA exhibit square lattices corresponding to the MgO(002) and CFA(002) crystal lattice planes when the CFA(002) X-ray diffraction vector (corresponds to the B2 phase) is perpendicular to film plane. On the other hand, CFA exhibits a hexagonal lattice and MgO a square lattice corresponding to the CFA(111) and MgO(002) crystal lattice planes, respectively when the CFA(111) X-ray diffraction vector (corresponds to the L2$_1$ phase) is perpendicular to film plane as indicated in Figure 1(a). A cross-sectional view of the CFA(8 nm) sample along the MgO [100] substrate direction is shown in Figure 1(b), where point 1 (hexagonal orientation) and point 2 (square orientation) correspond to the CFA(111) and CFA(002) crystal lattice planes, which confirms our supposition that the CFA film exhibits a mixed L2$_1$ and B2 phase up to 3 nm thickness. The matching of the Pt and CFA layers to the MgO buffer and substrate also confirms that our STEs are highly oriented. As CFA is known for its low damping, therefore, to further confirm our prediction, we have performed ferromagnetic resonance spectroscopy to obtain the intrinsic Gilbert damping parameter, which is found to be 5.58 (47) $\times$ 10$^{-3}$, which again indicates the mixed L2$_1$ and B2 phases of CFA. More details can be found in supplementary information section III.

\begin{figure*}[t]
    \centering
    \includegraphics[width=15cm]{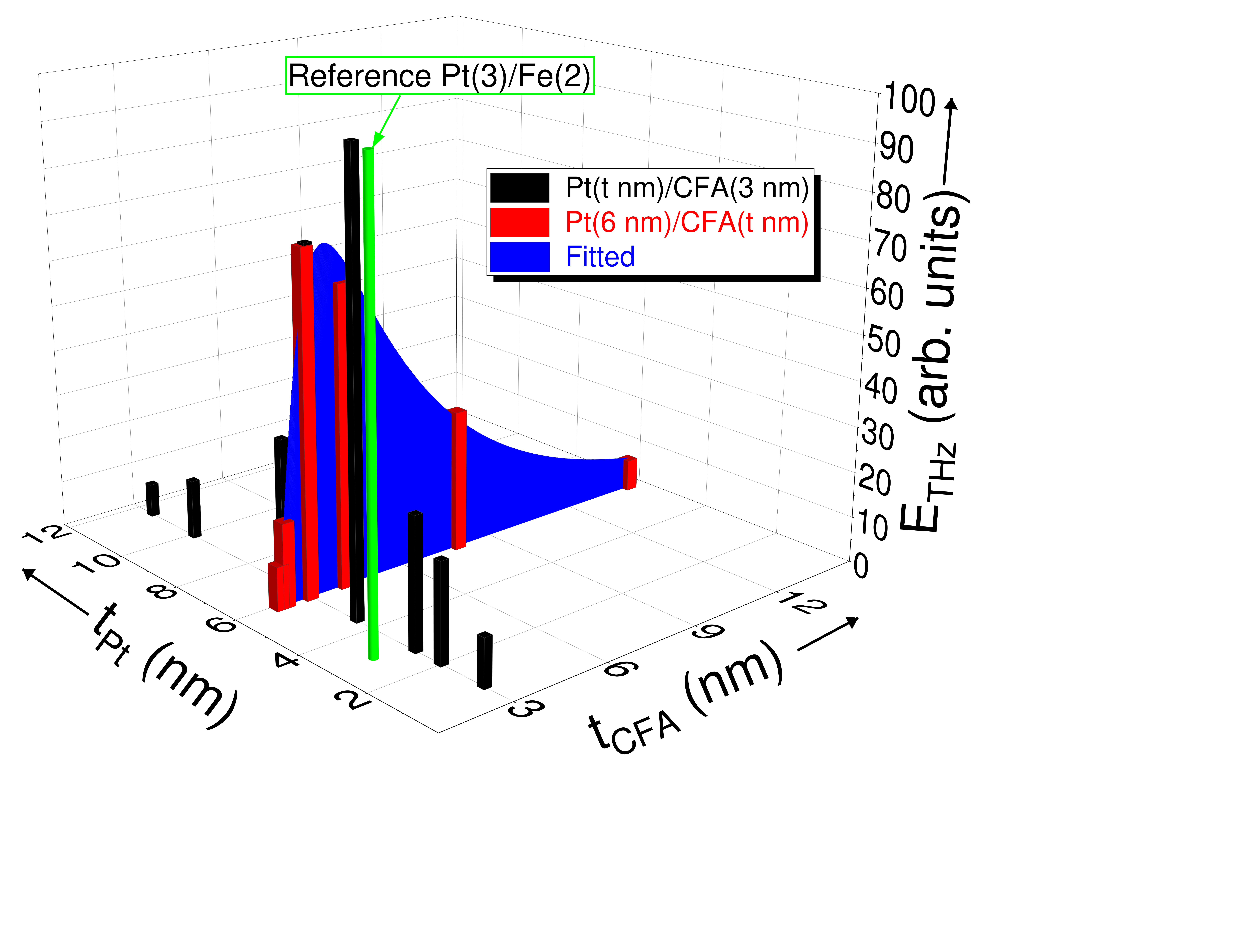}
    \caption{CFA (red) and Pt (black) thickness dependent THz electric field peak-to-peak amplitude. The blue curve is fitted with a CFA thickness dependent THz electric field amplitude using equation (3).  Green color corresponds to THz electric field peak-to-peak amplitude for the Pt(3)/Fe(2)/MgO STE.}
    \label{fig:my_label}
\end{figure*}

To compare the performance of our CFA/Pt STEs, we have also deposited the Fe(2 nm)/Pt(3 nm) reference sample on MgO substrate.
To optimize the parameters of the CFA/Pt STE, we utilized the THz-TDS set-up schematically shown in Figure 2(a) (cf. supplementary information section VI). The THz emission along with its fast Fourier transform spectrum for the reference Pt(3)/Fe(2)/MgO, Pt(4.5)/CFA(3)/MgO(10)/MgO, and Pt(6)/CFA(3)/MgO(10)/MgO STEs are shown in Figures 2(b,c). The peak-to-peak amplitude of THz E-field is found to be similar for the  Pt(3)/Fe(2)/MgO and Pt(4.5)/CFA(3)/MgO(10)/MgO STEs. Moreover, there is some ringing in the time domain signal after the main THz pulse due to multi-order reflections of the THz pulse within the sample, originating at the interface between the substrate and Si-lens and back side of the substrate (i.e., laser pump side). The effect of the ringing is also visible in the FFT of the signal as shown in Figure (2c). This effect of the ringing can be seen in several publications displaying the THz signal in the time and frequency domains \cite{klier2015influence,kinoshita2016terahertz,torosyan2018optimized,nandi2019antenna}. The peak-to-peak amplitude of the THz signal ($E_{THz}$) in time-domain is measured for all CFA/Pt STEs and used as the dependent variable in equation (3) with the CFA thickness as an independent variable. The explanation of equation (3) is given in the following section. 

A relation between the THz electric field and the charge current \textbf{($j_c=-e \int_{0}^t \gamma(z) j_s(z,\omega)dz$)} in the NFM layer can be formulated using Ohm's law by assuming that the electromagnetic radiation propagates as a plane-wave perpendicular to the STE film plane (\textit{i.e.,} z-direction as shown in Figure 2a), as the sample thickness is much smaller then the wavelength of the THz radiation. The THz electric field after the STE is described by $E_{THz} = Z(\omega)$ \\
$e \int_{0}^t \gamma(z) j_s(z,\omega)dz$, where $e$ is the charge of the electron, $t$ ($=t_{CFA}$ + $t_{Pt}$) is the total thickness of the STE, $\gamma(z)$ is the spin to charge conversion efficiency, and $j_s$ is the spin current density. $Z(\omega)$ is the impedance, which describes the conversion efficiency of charge current into THz electric field in the NFM layer, that can be expressed as \cite{seifert2018spintronics},
\begin{equation}
    \frac{1}{Z(\omega)} = \frac{(n_{air}+n_{MgO})}{Z_0} + \int_{0}^t \sigma(z,\omega)dz,
\end{equation}
where $\omega$/2$\pi$ is the frequency of the THz radiation, $n_{air}$ and $n_{MgO}$ are the refractive index of air and the MgO substrate, respectively, $Z_0$ ($\sim377$ $\Omega$) is the impedance of free space, and $\sigma$ is the electrical conductivity of the metal layers.

Several mechanisms have been proposed for describing generation of the optically induced spin polarised current in the FM layer, such as a superdiffusive spin current due to spin filtering of hot electrons \cite{PhysRevLett.105.027203}, the spin-Seebeck effect \cite{choi2015thermal}, and a demagnetization process based on  magnon-phonon coupling \cite{choi2014spin}. Since the THz electric field amplitude is dependent on the spin polarised current generated in the CFA layer, one can expect a larger amplitude due to multiple reflections of the pump pulse and therefore the Pt/CFA heterostructure can be treated as an Fabry-Perot cavity. The shorter the cavity volume is in the transverse direction (\textit{i.e.,} z-direction as shown in Figure 2a), the more echos occur, which may cause an enhancement of the spin current, and as a result the THz amplitude increases. However, below a certain thickness of the FM layer, the possibility of having an out-of-plane component of the spin polarization in the FM layer or even a superparamagnetic behaviour in case of island growth cannot be ruled out, which will cause a decrease of the THz electric field amplitude as indicated for the minimum CFA thickness used in this study (cf. Figure 3)). Therefore, due to this assumption \cite{PhysRevLett.69.3385,torosyan2018optimized}, $E_{THz}$ is proportional to $\tanh{[\frac{t_{CFA}-t_{c}}{2\lambda_{pol}}]}$, where $t_c$ is the critical layer thickness and  $\lambda_{pol}$ is a characteristic length that describes the saturation of the spin current when the thickness of the FM layer is larger than the critical thickness. $t_c$ is the thickness where the magnetization has an out-of-plane component, and therefore the applied magnetic field is not sufficient to saturate all the magnetic spins in film plane direction, as a result the lowest amplitude of THz-electric field is found as shown in Figure (3). Thickness dependent magnetization measurements are discussed in the forthcoming section.

\begin{figure*}[t]
    \centering
    \includegraphics[width=20cm]{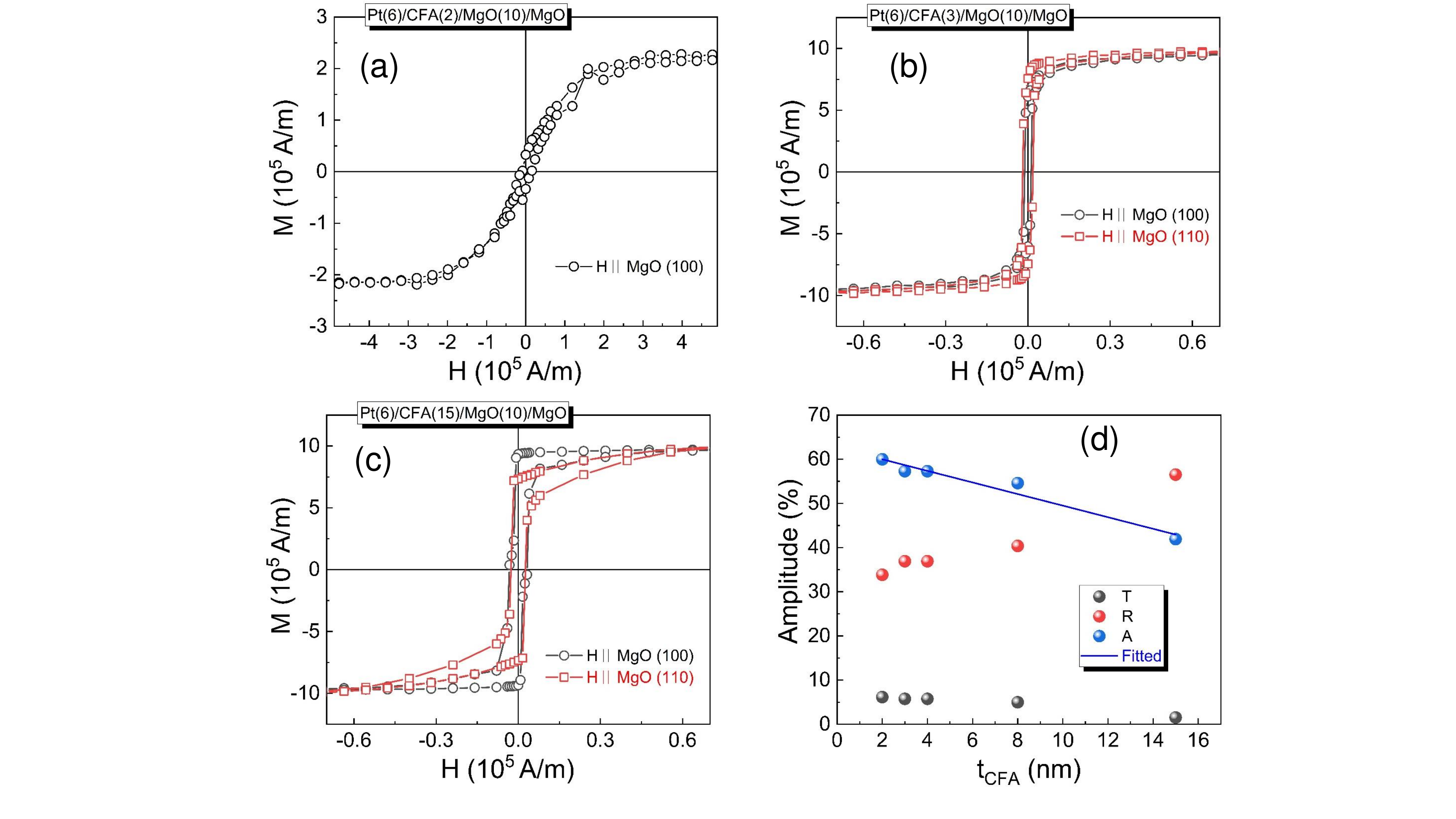}
    \caption{Magnetization versus applied magnetic field for (a) Pt(6)/CFA(2)/MgO(10)/MgO, (b) Pt(6)/CFA(3)/MgO(10)/MgO, and (c) Pt(6)/CFA(15)/MgO(10)/MgO samples. (d) Tranmittance ($T$), reflectance ($R$) and absorbance ($A$) of the CFA/Pt STEs. The laser wavelength was 800 nm.}
    \label{fig:my_label}
\end{figure*}
The generated spin current super diffuses from CFA to Pt through the CFA/Pt interface. Considering reflection of the spin current at the interface, the spatial dependence of the spin polarised current inside in the Pt layer is defined as \cite{PhysRevLett.104.046601}, 
\begin{equation}
    j_s (z)=j_s^0 (CFA) \frac{\sinh((z-t_{Pt})/\lambda_{Pt})}{\sinh(t_{Pt}/\lambda_{Pt})},
\end{equation}
where $j_s^0(CFA)$ is the spin current density at the CFA/Pt interface and $\lambda_{Pt}$ is the spin diffusion length in Pt. The emitted THz electric field is linearly dependent on the pump fluency, which indicates that the spin current is proportional to the energy density of the pump pulse. The measurements were performed at $\sim$40 mW laser pump power corresponding to a pump fluency of about 0.15 mJ/cm$^2$ calculated for an estimated laser spot size of 20 $\mu$m in diameter. This power level is well below the region, starting at about 1mJ/cm$^2$, where the THz signal starts to deviate from a linear relationship between THz power and the laser pump fluency. The external magnetic field was fixed at 85 mT (0.67 $\times$ 10$^5$ A/m) during the measurements. Therefore, $j_s^0(CFA)$ will scale as $A/t$, where $A$ is the absorbance of the incident pump pulse by the STE.
To know the absorbance, we have measured the reflected ($P_{ref}$) and transmitted ($P_{trans}$) powers in the optical region as a function of CFA thickness and calculated using $A=1-T-R$ as shown in Figure 4(d), where $T$ ($=P_{trans}/P_{inc}$) and $R$ ($=P_{ref}/P_{inc}$) are the  transmittance and reflectance in the optical region, and $P_{inc}$ is the incident power. The absorbance for all STEs was found to be larger than  $\sim$50\%, following the trend of the inverse of the CFA thickness; the calculated absorbance is used as input parameter in equation (3). 

As we have described the electromagnetic field as a plane wave inside the CFA/Pt STE there is no loss considered in the Green's function approach \cite{kampfrath2013terahertz}. However, taking a small loss into account, the propagating vector of the plane wave will have a complex form, resulting in an additional attenuation factor $\exp^{-t/\zeta_{THz}}$ for the THz electric field, where $\zeta_{THz}$ is the effective inverse attenuation coefficient. This factor is only dependent upon the total thickness of the layers \cite{torosyan2018optimized}. Therefore, making use of all assumptions mentioned above in Ohm's law, the final form of the THz electric field can be expressed as,

\begin{eqnarray}
    E_{THz} \propto  \frac{A}{t} \tanh\left(\frac{t_{CFA}-t_{c}}{2\lambda_{pol}}\right) \cdot \tanh\left(\frac{t_{Pt}}{2\lambda_{Pt}}\right) \\ \nonumber
    \cdot \frac{1}{n_{air}+n_{MgO}+Z_0(\sigma_{CFA}t_{CFA}+\sigma_{Pt}t_{Pt})}e^{-t/\zeta_{THz}},
\end{eqnarray}\\
where $\sigma_{CFA}$ and $\sigma_{Pt}$ are the electrical conductivities of CFA and Pt, respectively. The experimental data for the THz electric field have been fitted using equation (3) as shown in Figure 3. We have used the experimentally measured $A\sim$50\%, $n_{air}$=1, $n_{MgO}$=3.1 (cf. supplementary information section VII), $\sigma_{CFA} = 1.414$ MS/m (cf. supplementary information section V), $\sigma_{Pt} = 2$ MS/m, $\lambda_{Pt} = 1.4$ nm \cite{torosyan2018optimized} as fixed parameters and $t_c$, $\lambda_{pol}$ and $\zeta_{THz}$ as global fitting parameters, yielding $t_c = 1.98 (09)$ nm, $\lambda_{pol}= 0.209 (89)$ nm and $\zeta_{THz} = 20.4 (16)$ nm for the CFA/Pt STE. 

To understand the THz electric field amplitude below the critical thickness, we have studied the thickness (2, 3 and 15 nm) dependent static magnetic properties of the CFA films capped with Pt(6nm). It can be seen that the samples with 3 and 15 nm thickness exhibit in-plane magnetization with square-like hysteresis loops following  the expected behavior of epitaxially/highly oriented grown CFA thin films \cite{PhysRevB.87.184431}. However, in case of the 2nm thin CFA film, the magnetization saturates at a relatively higher magnetic field as shown in Figure 4a. Moreover, the hysteresis curve is not square-like, which indicates that the magnetization is dominated by an out-of-plane or easy-cone anisotropy. Thus, the applied magnetic field used in the THz measurements is not sufficient to saturate the magnetization, which results into a low value for the THz electric field amplitude. 

In the above model (equation 3), we did not consider the effect of the Si lens, which is used to focus the emitted THz radiation towards the detector. Moreover, the amplitude of the THz electric field depends upon the interface roughness; the higher the interface roughness, the lower the THz electric field amplitude \cite{PhysRevMaterials.3.084415}. This may be due to interfacial spin memory loss (SML) at the FM/Pt interface as the SML depends on the interface mixing/roughness \cite{PhysRevLett.124.087702}. The roughness of our CFA/Pt STEs lies in the range of 0.5-1.1 nm. It means that there is still room for improvements with respect to the THz electric field amplitude by reducing the interface roughness of the CFA/Pt STE. Importantly, the bandwidth is found to be 0.1-4 THz for CFA/Pt STEs, which is limited by the low temperature (LT)-GaAs detector used in our system. By changing the LT-GaAs detector to a ZnTe detector and using a shorter pulse width of the femtosecond laser, we would expect a bandwidth of our STEs upto 30 THz \cite{Seifert2016efficient}.

In summary, the Co$_2$FeAl full Heusler compound based spintronic terahertz emitter (STE) is reported, which has a bandwidth in the range of 0.1-4 THz. We have deposited an ordered L2$_1$ and B2 mixed phase of Co$_2$FeAl in the ultrathin film regime at room temperature using MgO(10 nm) as seed layer. We utilized the thickness-dependent THz electric field model to extract the THz parameters for our Co$_2$FeAl/Pt STEs. The THz electric field amplitude was optimized with respect to the thickness, growth parameters, and orientation. The THz electric field is found to be maximum for Pt(6)/CFA(3) STE and minimum for Pt(6)/CFA(2) STE, respectively. Our study reveals that below 3 nm CFA thickness, the THz electric field amplitude  decreases, which is explained by the out-of-plane component of the magnetization. This study provides a unique direction for STE research, utilizing the Co-based full Heusler compounds.

\begin{acknowledgments}
This work is supported by the Swedish Research Council (grant no. 2017–03799) and Olle Engkvists Stiftelse (grant no. 182–0365). RG acknowledges Vassilios Kapaklis for help in the XRD measurments and Lars Rieker for TEM measurements. 

\end{acknowledgments}
\nocite{*}

\bibliography{apssamp}% Produces the bibliography via BibTeX.
\bibliographystyle{apsrev4-2}
%\printbibliography

\end{document}